\documentclass[conference]{IEEEtran}
\IEEEoverridecommandlockouts
\usepackage[utf8]{inputenc}
\usepackage{cite}
\usepackage{lscape}
\usepackage{textcomp}
\usepackage{multirow}
\usepackage[graphicx]{realboxes}
\usepackage{adjustbox}
\usepackage{tabularx}
\usepackage{amsmath}
\usepackage{amssymb}
\usepackage{multirow, makecell}
\usepackage{array}
\usepackage{adjustbox}
\usepackage{tablefootnote}
\usepackage{verbatim}
\usepackage{graphicx}

\def\BibTeX{{\rm B\kern-.05em{\sc i\kern-.025em b}\kern-.08em
    T\kern-.1667em\lower.7ex\hbox{E}\kern-.125emX}}
\begin{document}
\title{Decoding of Intuitive Visual Motion Imagery Using Convolutional Neural Network under 3D-BCI Training Environment
\footnote{{\thanks{\hrule Research was partly supported by Institute of Information \& Communications Technology Planning \& Evaluation (IITP) grant funded by the Korea government (No. 2017-0-00432, Development of Non-Invasive Integrated BCI SW Platform to Control Home Appliances and External Devices by User’s Thought via AR/VR Interface) and partly funded by Institute of Information \& Communications Technology Planning \& Evaluation (IITP) grant funded by the Korea government (No. 2017-0-00451, Development of BCI based Brain and Cognitive Computing Technology for Recognizing User’s Intentions using Deep Learning).}
\thanks{© 20xx IEEE. Personal use of this material is permitted. Permission
from IEEE must be obtained for all other uses, in any current or future media, including reprinting/republishing this material for advertising or promotional purposes, creating new collective works, for resale or redistribution to servers or lists, or reuse of any copyrighted component of this work in other works.}}
}}

\author{\IEEEauthorblockN{Byoung-Hee Kwon}
\IEEEauthorblockA{\textit{Dept. Brain and Cognitive Engineering}\\
\textit{Korea University} \\
Seoul, Republic of Korea \\
kwonbh1990@korea.ac.kr}\\

\IEEEauthorblockN{Jeong-Hyun Cho}
\IEEEauthorblockA{\textit{Dept. Brain and Cognitive Engineering}\\
\textit{Korea University} \\
Seoul, Republic of Korea \\
jh\_cho@korea.ac.kr}

\and

\IEEEauthorblockN{Ji-Hoon Jeong}
\IEEEauthorblockA{\textit{Dept. Brain and Cognitive Engineering}\\
\textit{Korea University} \\
Seoul, Republic of Korea \\
jh\_jeong@korea.ac.kr}\\

\IEEEauthorblockN{Seong-Whan Lee}
\IEEEauthorblockA{\textit{Dept. Artificial Intelligence}\\
\textit{Korea University} \\
Seoul, Republic of Korea \\
sw.lee@korea.ac.kr}
}

\maketitle

\begin{abstract}
In this study, we adopted visual motion imagery, which is a more intuitive brain-computer interface (BCI) paradigm, for decoding the intuitive user intention. We developed a 3-dimensional BCI training platform and applied it to assist the user in performing more intuitive imagination in the visual motion imagery experiment. The experimental tasks were selected based on the movements that we commonly used in daily life, such as picking up a phone, opening a door, eating food, and pouring water. Nine subjects participated in our experiment. We presented statistical evidence that visual motion imagery has a high correlation from the prefrontal and occipital lobes. In addition, we selected the most appropriate electroencephalography channels using a functional connectivity approach for visual motion imagery decoding and proposed a convolutional neural network architecture for classification. As a result, the averaged classification performance of the proposed architecture for 4 classes from 16 channels was 67.50 ($\pm$1.52)$\%$ across all subjects. This result is encouraging, and it shows the possibility of developing a BCI-based device control system for practical applications such as neuroprosthesis and a robotic arm.\\

\end{abstract}


\section{Introduction}
Brain-computer interface (BCI) is the technology that creates communication pathways between users and computers possible using brain signals \cite{vaughan2003brain}, \cite{abiri2018comprehensive}. Electroencephalography (EEG) has an advantage of higher time resolution than comparable methods such as functional magnetic resonance imaging and near-infrared spectroscopy \cite{ zhu2016canonical, ding2013changes}. Therefore, EEG-based BCI enables instant communication between users and computers based on the advantages of the higher time resolution.

\begin{figure*}[t!]
\centering
\centerline{\includegraphics[width=\textwidth]{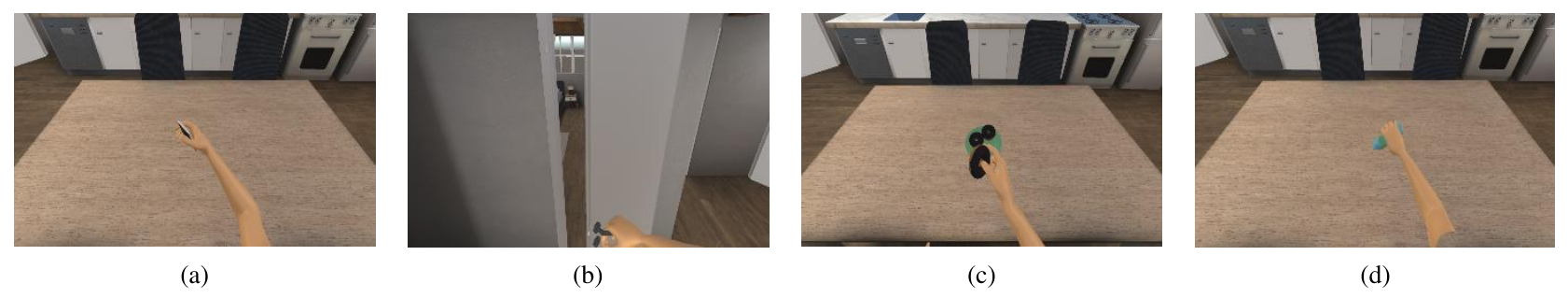}}
\caption{3D virtual BCI platform-based stimuli given to users in experiment. Each stimulus represents the movement frequently used in real life in a three-dimensional space. Each stimulus consists of (a) picking up a cell phone, (b) opening a door, (c) eating food, and (d) pouring water.}
\label{fig:1}
\end{figure*}

The BCI system controls the device according to user's intention\cite{wolpaw2002brain}. Many related studies for decoding the user's movement intentions with sufficient accuracy were performed using various BCI paradigms \cite{jeong2020decoding, antelis2016decoding}. For example, many BCI investigators have adopted a variety of paradigms such as steady-state visual evoked potential (SSVEP) \cite{kwak2015lower, benda2020comparison, chen2016high} and event-related potential (ERP) \cite{fazel2012p300, li2018hybrid}. Among the BCI paradigms, motor imagery (MI) \cite{ang2017eeg} is one of the effective options to develop an intuitive BCI system for controlling practical applications such as robotic arm \cite{hochberg2012reach, jeong2020Brain, meng2018three} or wheelchair \cite{kim2018commanding}. Using exogenous paradigms (e.g., ERP and SSVEP) require the additional stimulation devices including the BCI system itself. This inconvenience limits the user from controlling devices using BCI with a concentration in the real-world environment. To overcome the limitation, many BCI investigators rely on endogenous paradigms such as MI because it does not require the external stimulation \cite{hochberg2012reach, lee2015subject, edelman2019noninvasive}. MI is one of the endogenous paradigms similar to the visual motion imagery which is used for this study, but it has disadvantages that are difficult to imagine intuitively. In the MI paradigm, the user should perform imagery of the movement on their muscles. However, this kind of imagery is perceived differently by each user therefore it causes a lack of uniformity. Even if the user performs the imagery of the same task, motor imagery relies on the imagination of muscle movement which is different between each user affects the activation region of the brain and the activation of the brain, and triggers different brain signals for each user. In order to control the device in the online environment based on the BCI, it is necessary to calibrate for a long time \cite{jeong2020eeg}. At this time, MI causes user fatigue due to the difficulty of the imagery task, and it reduces BCI performance. To overcome this limitation, in this study, we adopted the visual motion imagery as the BCI paradigm.

Visual motion imagery allows users imagine more intuitively than MI when performing imagery tasks. Intuitive imagination enhances the user's brain signal quality and induces less fatigue when performing imagery tasks. Visual motion imagery can be analyzed in various frequency regions such as delta, theta, and alpha band, and the prefrontal and occipital regions are mainly activated \cite{koizumi2018development}. Brain activities based on visual motion imagery induce delta band in the prefrontal lobe and the alpha band in the occipital lobe. In previous studies, it was proved that the brain signal by visual motion imagery can be analyzed by frequency and region \cite{sousa2017pure}, and based on this, it can be applied to BCI to control the device.

In this study, we designed a virtual BCI training platform for visual motion imagery decoding which is a kind of intuitive imagination. The BCI training platform was constructed by applying virtual reality in three dimensions for more intuitive stimulation and imagination than previous visual motion imagery related research. The 3D BCI training platform consists of four movements, which are based on the movements needed for real life. Based on this stimulus, the user could perform a visual motion imagery task more easily. In addition, we adopted the deep learning approach based on the EEG functional connectivity for robust user intention decoding. Hence, we confirmed that the possibility of controlling a device such as a robotic arm in a real environment by decoding brain signals. In addition, this study will be valuable as research towards the intuitive visual motion imagery decoding.

\begin{figure}[t!]
\centering
\centerline{\includegraphics[width=0.8\columnwidth]{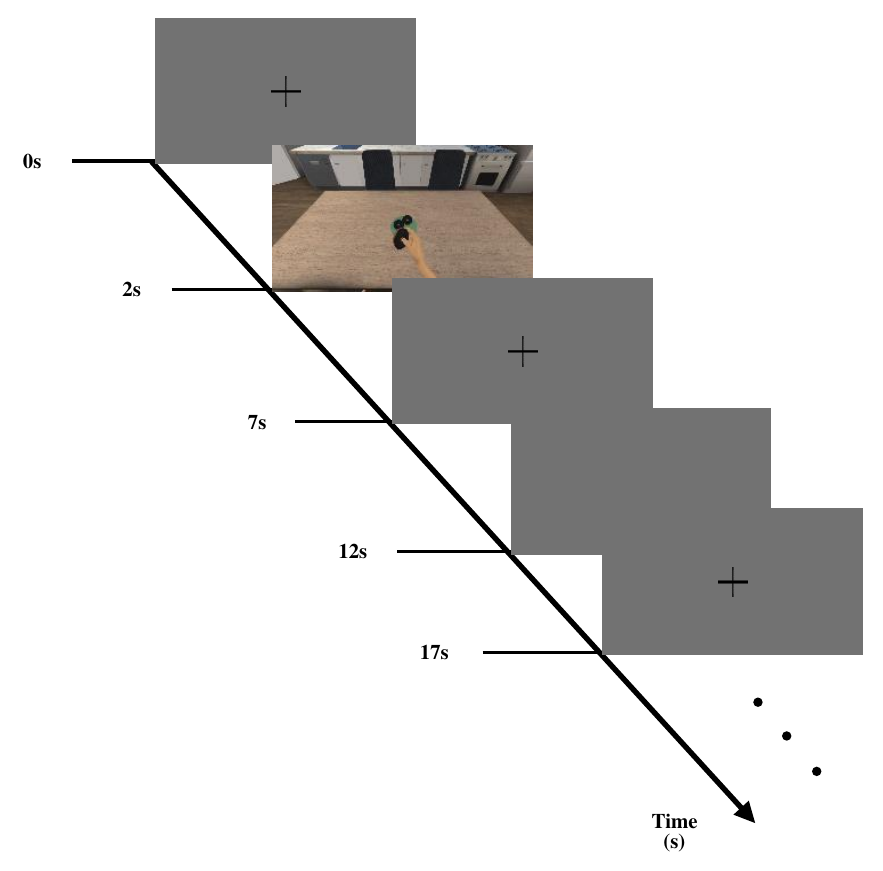}}
\caption{The overview of experimental paradigm. Overall the experiment consists of 200 trials with a total of 50 trials per class.}
\label{fig:2}
\end{figure}   

\section{Methods}

\subsection{Subjects}
Nine healthy subjects with no history of neurological disease were recruited for this experiment (S1--S9; ages 24-30; 5 men and 4 women, all right-handed). All subjects were naïve to the BCI experiment. Subjects were advised to avoid anything could affect physical and mental condition, such as drinking alcohol or having psychotropic drugs, and had enough sleep for more than 8 hours before the experiment day. This study was reviewed and approved by the Institutional Review Board at Korea University [KUIRB-2020-0013-01], and written informed consent was obtained from all subjects before the experiment.

\subsection{Data Acquisition}
EEG data were collected by 1,000 Hz sampling rate using 64 Ag/AgCl electrodes (Fp1--2, AF3--4, AF7--8, AFz, F1--8, Fz, FC1--6, FT7--10, C1--6, Cz, T7--8, CP1--6, CPz, TP7--10, P1--8, Pz, PO3--4, PO7--8, POz, O1--2, Oz, Iz) in the 10/20 international system via BrainAmp (BrainProduct GmbH, Germany). The impedance of all EEG electrodes was maintained below 10 k$\Omega$. The entire process of the experiment was performed in a shielded space from the light and noise so the subjects could concentrate on the experiment with seating in a comfortable chair.

\subsection{Experimental Protocols}
In the experiment, we designed the visual stimuli as the instructions in the experimental protocol, which gives guidance on what the subject should imagine. Following the given visual cue, the subject imagined the designated task. The videos that represent the designated tasks were created by 3-dimensional simulation software such as Unity 3D and Blender (Blender 3D Engine: www.blender.org). We prepared four different scenarios for visual stimuli, and those scenarios are opening a door, picking up a cell phone, eating food, and pouring water, respectively, as shown in Fig. 1. We expected that using these kinds of visual stimuli which is moving video instead of a still image allows the subjects to perform the visual motion imagery easier because the designated four classes are intuitive movement imagery in an upper limb for handling daily objects. The classes were consisted based on the movements that the user can actually use in real life.

To maintain the quality of EEG signals while the subjects performing visual motion imagery, we designed the experimental paradigm including four different phases, as described in Fig. 2. The first phase is a resting state for 2 s. It gives a comfortable environment for the subjects before performing visual motion imagery. The second phase is presenting a visual cue that gives a visual stimulus to the subjects about what class they have to imagine. The third phase is another resting state for 5 s to remove the afterimage and effect of the given visual stimulus in the previous stage.

Once the subjects had the 5-s length resting phase, they were asked to perform corresponding visual motion imagery for 5 s. During the visual motion imagery phase, the subjects were asked to see a blank screen with the eyes open and to imagine the visual stimulus seen in the previous phase like drawing a picture. Therefore, a single trial is 17 s long including all the four phases mentioned above, and each class consists of 50 trials. As a result, a total of 200 trials were collected for every subject.

\subsection{Data Analysis}
Data analysis was conducted using the BBCI Toolbox \cite{blankertz2010berlin} and EEGLAB Toolbox (version 14.1.2b) \cite{delorme2004eeglab} in an offline analysis. The EEGLAB toolbox was used to measure the variation of spectral power of the visual motion imagery at each frequency based on the event-related spectral perturbation (ERSP). ERSP analysis was performed between 3 and 50 Hz, using 400 time-points. We set a baseline from -500 to 0 ms before the start of the imagination (last 500 ms of rest phase) in order to analyze the phenomena when subjects imagining.

We performed a statistical analysis to compare the imagery phase and the resting phase of each class to prove that the experiment proceeded to visual motion imagery. For analysis, the raw EEG signal was band-pass filtered between [0.5--13] Hz including the delta, theta, and alpha band frequency generated in the visual motion imagery. The EEG signals were down-sampled from 1,000 to 250 Hz. Based on the procedure, we calculated power spectral density (PSD) for the four classes and the resting state. In addition, we performed a non-parametric paired permutation test for statistical analysis \cite{maris2007nonparametric} of visual motion imagery and confirmed the significance between the imagery phase of each class and the resting state. We applied statistical method to explore the spatial differences between visual motion imagery phase and resting phase in delta, theta, and alpha power.

\begin{figure}[t!]
\centering
\centerline{\includegraphics[width=\columnwidth]{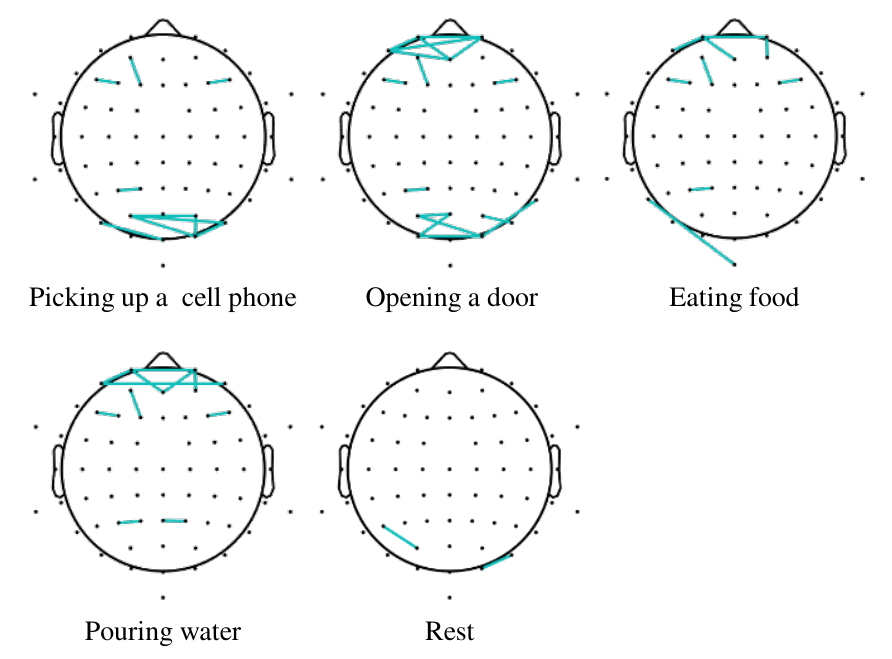}}
\caption{Channel connections selected by functional connectivity that have a correlation above 0.9. The connections between channels appear primarily in the prefrontal and occipital lobe.}
\label{fig:3}
\end{figure}

\begin{table*}[t!]
\caption{Design of The Proposed CNN Architecture}
\begin{center}
\small
\renewcommand{\arraystretch}{1.2}
\begin{tabular*}{\textwidth}{@{\extracolsep{\fill}\quad}ccccc}
\hline
Layer & Type            & Output shape                 & Kernel size       & Stride \\ \hline
1     & Convolution     & 1 × 25 × N of channels × 376 & 1 × 125            & 1 × 1  \\
2     & Convolution     & 1 × 25 × 1 × 376             & N of channels × 1 & 1 × 1  \\
3     & Average Pooling & 1 × 25 × 1 × 94             & 1 × 4             & 1 × 4  \\
4     & Convolution     & 1 × 50 × 1 × 80              & 1 × 15            & 1 × 1  \\
5     & Average Pooling & 1 × 50 × 1 × 20              & 1 × 4             & 1 × 4  \\
6     & Convolution     & 1 × 100 × 1 × 6             & 1 × 15            & 1 × 1  \\
7     & Average Pooling & 1 × 100 × 1 × 1              & 1 × 4             & 1 × 4  \\
8     & Flatten         & 1 × 100                      & -                 & -      \\
9     & SoftMax         & 1 × 4                        & -                 & -    \\\hline
\end{tabular*}
\end{center}
\end{table*}

\begin{figure*}[t!]
\centering
\centerline{\includegraphics[scale = 1.2, height=185pt]{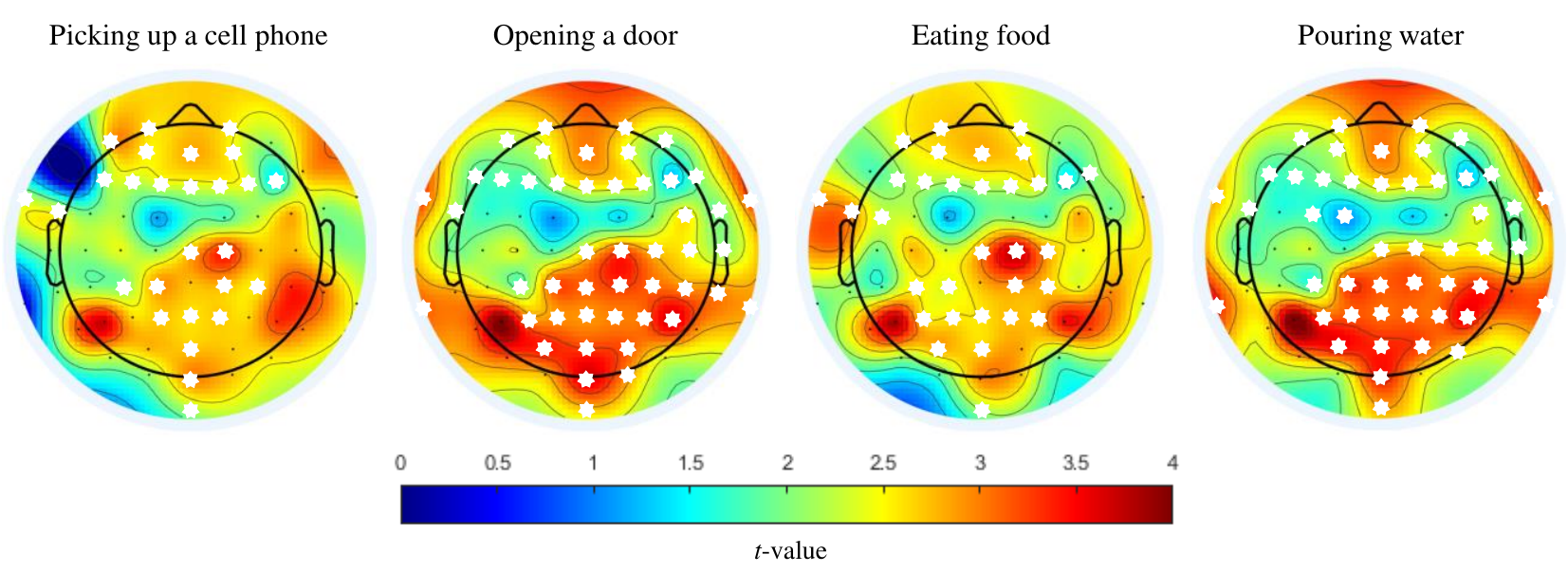}}
\caption{The results of statistical analysis using permutation test. The intensity of brain activation was expressed as \textit{t}-value. A white asterisk indicates an electrode that is significantly different the imagery phase compared to the resting phase (\textit{p} $\leq$ 0.01).}
\label{fig:4}
\end{figure*}

The critical issue in this study is that EEG-based functional connectivity \cite{ding2013changes} to find visual motion imagery-related channels using the correlation between channels for analyzing spatial information. The EEG signals were pre-processed to be suitable for the functional connectivity analysis. Using the pre-processed EEG signals, we applied trial-averaged phase locking value method to measure the connectivity between channels. Fig. 3 shows the functional connectivity corresponding to each class. For the functional connectivity in the visual motion imagery, channels with correlation mainly in the prefrontal and occipital lobes, which have the values of above 0.9 were particularly connected to each other. Therefore, we proposed the channel selection method base on the result of the statistical analysis of visual motion imagery. With the proposed method, we could select the specific channels associated with the visual motion imagery and use the selected channels layout to configure the BCI-based device with fewer channels.
 
\subsection{Architecture Design}
 Initially, we conducted the channel selection in the EEG pre-processing step. Since we extracted spatial information based on the functional connectivity approach, and we designed the convolutional neural network (CNN) architecture for visual motion imagery decoding as depicted in Table I.
 
 The proposed architecture is composed of 4 convolution layers and 3 pooling layers. In order to obtain temporal information of EEG, it is necessary to increase the number of convolution layers. But due to the limitation of the number of training sets, using more layers can cause a decrease in performance. Batch normalization was applied to normalize the variance of learned data, and a dropout layer was applied to prevent overfitting problems. The EEG data was band-pass filtered in the frequency range between 0.5 and 13 Hz, which is related to the visual motion imagery, and analyzed from 500 to 4500 ms, which is the time when it is judged to be actually imagining. 
 
 In addition, for better deep learning performance, the sliding window was performed with an augmentation technique at 50\% overlapping every 2 s. We used CSP--LDA \cite{ang2012filter} and deepConvNet \cite{schirrmeister2017deep} as the control groups to compare the classification performance of the proposed architecture.
 
\begin{table*}[t!]
\caption{Classification Performance of Proposed Architecture According to Each Channel Group}
\begin{center}
\small
\renewcommand{\arraystretch}{1.15}
\begin{tabular*}{\textwidth}{@{\extracolsep{\fill}\quad}ccccccccccc}
\hline
Channel & Sub 1                                                     & Sub 2                                                     & Sub 3                                                     & Sub 4                                                     & Sub 5                                                     & Sub 6                                                     & Sub 7                                                     & Sub 8                                                     & Sub 9                                                     & Avgerage                                                  \\ \hline
2ch     & \begin{tabular}[c]{@{}c@{}}75.63\%\\ (±1.25)\end{tabular} & \begin{tabular}[c]{@{}c@{}}42.50\%\\ (±2.21)\end{tabular} & \begin{tabular}[c]{@{}c@{}}58.75\%\\ (±2.78)\end{tabular} & \begin{tabular}[c]{@{}c@{}}54.38\%\\ (±2.55)\end{tabular} & \begin{tabular}[c]{@{}c@{}}42.50\%\\ (±1.78)\end{tabular} & \begin{tabular}[c]{@{}c@{}}58.75\%\\ (±1.35)\end{tabular} & \begin{tabular}[c]{@{}c@{}}70.01\%\\ (±2.43)\end{tabular} & \begin{tabular}[c]{@{}c@{}}31.88\%\\ (±1.97)\end{tabular} & \begin{tabular}[c]{@{}c@{}}58.13\%\\ (±3.10)\end{tabular} & \begin{tabular}[c]{@{}c@{}}54.72\%\\ (±2.33)\end{tabular} \\
4ch     & \begin{tabular}[c]{@{}c@{}}74.38\%\\ (±3.51)\end{tabular} & \begin{tabular}[c]{@{}c@{}}41.25\%\\ (±2.75)\end{tabular} & \begin{tabular}[c]{@{}c@{}}58.13\%\\ (±1.38)\end{tabular} & \begin{tabular}[c]{@{}c@{}}52.50\%\\ (±0.75)\end{tabular} & \begin{tabular}[c]{@{}c@{}}39.38\%\\ (±1.49)\end{tabular} & \begin{tabular}[c]{@{}c@{}}57.50\%\\ (±2.25)\end{tabular} & \begin{tabular}[c]{@{}c@{}}72.50\%\\ (±1.81)\end{tabular} & \begin{tabular}[c]{@{}c@{}}34.38\%\\ (±3.01)\end{tabular} & \begin{tabular}[c]{@{}c@{}}64.38\%\\ (±1.20)\end{tabular} & \begin{tabular}[c]{@{}c@{}}54.93\%\\ (±2.57)\end{tabular} \\
8ch     & \begin{tabular}[c]{@{}c@{}}70.00\%\\ (±0.99)\end{tabular} & \begin{tabular}[c]{@{}c@{}}63.75\%\\ (±2.44)\end{tabular} & \begin{tabular}[c]{@{}c@{}}60.00\%\\ (±2.49)\end{tabular} & \begin{tabular}[c]{@{}c@{}}70.00\%\\ (±1.49)\end{tabular} & \begin{tabular}[c]{@{}c@{}}53.75\%\\ (±1.75)\end{tabular} & \begin{tabular}[c]{@{}c@{}}63.13\%\\ (±3.55)\end{tabular} & \begin{tabular}[c]{@{}c@{}}83.75\%\\ (±1.23)\end{tabular} & \begin{tabular}[c]{@{}c@{}}35.63\%\\ (±1.52)\end{tabular} & \begin{tabular}[c]{@{}c@{}}78.75\%\\ (±2.48)\end{tabular} & \begin{tabular}[c]{@{}c@{}}64.31\%\\ (±2.11)\end{tabular} \\
16ch    & \begin{tabular}[c]{@{}c@{}}81.88\%\\ (±1.48)\end{tabular} & \begin{tabular}[c]{@{}c@{}}72.50\%\\ (±2.71)\end{tabular} & \begin{tabular}[c]{@{}c@{}}62.50\%\\ (±2.69)\end{tabular} & \begin{tabular}[c]{@{}c@{}}65.63\%\\ (±2.11)\end{tabular} & \begin{tabular}[c]{@{}c@{}}61.88\%\\ (±0.95)\end{tabular} & \begin{tabular}[c]{@{}c@{}}61.25\%\\ (±1.14)\end{tabular} & \begin{tabular}[c]{@{}c@{}}85.00\%\\ (±1.37)\end{tabular} & \begin{tabular}[c]{@{}c@{}}41.25\%\\ (±1.52)\end{tabular} & \begin{tabular}[c]{@{}c@{}}75.63\%\\ (±1.30)\end{tabular} & \begin{tabular}[c]{@{}c@{}}67.50\%\\ (±1.52)\end{tabular} \\
20ch    & \begin{tabular}[c]{@{}c@{}}81.88\%\\ (±1.71)\end{tabular} & \begin{tabular}[c]{@{}c@{}}66.88\%\\ (±2.01)\end{tabular} & \begin{tabular}[c]{@{}c@{}}57.50\%\\ (±2.44)\end{tabular} & \begin{tabular}[c]{@{}c@{}}64.38\%\\ (±1.98)\end{tabular} & \begin{tabular}[c]{@{}c@{}}51.25\%\\ (±2.17)\end{tabular} & \begin{tabular}[c]{@{}c@{}}58.13\%\\ (±1.82)\end{tabular} & \begin{tabular}[c]{@{}c@{}}83.75\%\\ (±1.57)\end{tabular} & \begin{tabular}[c]{@{}c@{}}40.38\%\\ (±1.98)\end{tabular} & \begin{tabular}[c]{@{}c@{}}72.50\%\\ (±2.14)\end{tabular} & \begin{tabular}[c]{@{}c@{}}64.07\%\\ (±1.57)\end{tabular} \\
32ch    & \begin{tabular}[c]{@{}c@{}}82.50\%\\ (±1.53)\end{tabular} & \begin{tabular}[c]{@{}c@{}}58.13\%\\ (±2.30)\end{tabular} & \begin{tabular}[c]{@{}c@{}}63.13\%\\ (±2.41)\end{tabular} & \begin{tabular}[c]{@{}c@{}}62.50\%\\ (±1.88)\end{tabular} & \begin{tabular}[c]{@{}c@{}}55.63\%\\ (±1.41)\end{tabular} & \begin{tabular}[c]{@{}c@{}}62.50\%\\ (±1.87)\end{tabular} & \begin{tabular}[c]{@{}c@{}}77.50\%\\ (±1.58)\end{tabular} & \begin{tabular}[c]{@{}c@{}}40.63\%\\ (±1.92)\end{tabular} & \begin{tabular}[c]{@{}c@{}}70.63\%\\ (±2.01)\end{tabular} & \begin{tabular}[c]{@{}c@{}}63.68\%\\ (±1.88)\end{tabular} \\
64ch    & \begin{tabular}[c]{@{}c@{}}76.88\%\\ (±3.18)\end{tabular} & \begin{tabular}[c]{@{}c@{}}63.75\%\\ (±2.29)\end{tabular} & \begin{tabular}[c]{@{}c@{}}65.00\%\\ (±1.93)\end{tabular} & \begin{tabular}[c]{@{}c@{}}72.50\%\\ (±1.52)\end{tabular} & \begin{tabular}[c]{@{}c@{}}56.25\%\\ (±1.45)\end{tabular} & \begin{tabular}[c]{@{}c@{}}58.75\%\\ (±2.18)\end{tabular} & \begin{tabular}[c]{@{}c@{}}78.75\%\\ (±1.24)\end{tabular} & \begin{tabular}[c]{@{}c@{}}38.13\%\\ (±2.02)\end{tabular} & \begin{tabular}[c]{@{}c@{}}72.50\%\\ (±1.42)\end{tabular} & \begin{tabular}[c]{@{}c@{}}64.72\%\\ (±2.13)\end{tabular} \\ \hline
\end{tabular*}
\end{center}
\end{table*}

\begin{table}[t!]
\caption{Performance Comparison using the conventional methods}
\resizebox{\columnwidth}{!}{
\small
\renewcommand{\arraystretch}{1.3}
\begin{tabular}{@{\extracolsep{\fill}\quad}cccccccc}
\hline
Models       & 2ch                                                       & 4ch                                                       & 8ch                                                       & 16ch                                                      & 20ch                                                      & 32ch                                                      & 64ch                                                      \\ \hline
CSP--LDA \cite{ang2012filter}        & \begin{tabular}[c]{@{}c@{}}29.83\%\\ (±3.58)\end{tabular} & \begin{tabular}[c]{@{}c@{}}32.74\%\\ (±2.99)\end{tabular} & \begin{tabular}[c]{@{}c@{}}30.53\%\\ (±3.87)\end{tabular} & \begin{tabular}[c]{@{}c@{}}33.50\%\\ (±3.21)\end{tabular} & \begin{tabular}[c]{@{}c@{}}32.60\%\\ (±2.98)\end{tabular} & \begin{tabular}[c]{@{}c@{}}30.94\%\\ (±3.14)\end{tabular} & \begin{tabular}[c]{@{}c@{}}30.97\%\\ (±2.51)\end{tabular} \\
DeepConvNet \cite{schirrmeister2017deep} & \begin{tabular}[c]{@{}c@{}}52.08\%\\ (±2.75)\end{tabular} & \begin{tabular}[c]{@{}c@{}}54.35\%\\ (±1.98)\end{tabular} & \begin{tabular}[c]{@{}c@{}}60.42\%\\ (±2.21)\end{tabular} & \begin{tabular}[c]{@{}c@{}}62.29\%\\ (±1.54)\end{tabular} & \begin{tabular}[c]{@{}c@{}}61.99\%\\ (±1.33)\end{tabular} & \begin{tabular}[c]{@{}c@{}}61.18\%\\ (±2.14)\end{tabular} & \begin{tabular}[c]{@{}c@{}}63.47\%\\ (±2.68)\end{tabular} \\
\textbf{Proposed}    & \begin{tabular}[c]{@{}c@{}}\textbf{54.72\%}\\ \textbf{(±2.33)}\end{tabular} & \begin{tabular}[c]{@{}c@{}}\textbf{54.93\%}\\ \textbf{(±2.57)}\end{tabular} & \begin{tabular}[c]{@{}c@{}}\textbf{64.31\%}\\ \textbf{(±2.11)}\end{tabular} & \begin{tabular}[c]{@{}c@{}}\textbf{67.50\%}\\ \textbf{(±1.52)}\end{tabular} & \begin{tabular}[c]{@{}c@{}}\textbf{64.07\%}\\ \textbf{(±1.57)}\end{tabular} & \begin{tabular}[c]{@{}c@{}}\textbf{63.68\%}\\ \textbf{(±1.88)}\end{tabular} & \begin{tabular}[c]{@{}c@{}}\textbf{64.72\%}\\ \textbf{(±2.13)}\end{tabular} \\ \hline

\end{tabular}
}
\end{table}

\section{Results and Discussion}
\subsection{Statistical Analysis}
We investigated significant differences in the degree of brain activation of each class based on one versus rest approach. In specific, Fig. 4 showed the spatial differences in spectral power for each class and resting state using statistical analysis. The results indicated that all visual motion imagery is mainly activated in the prefrontal and occipital lobes. On the other hand, there was no statistically significant difference in other brain regions.

\begin{figure}[t!]
\centering
\centerline{\includegraphics[width=\columnwidth]{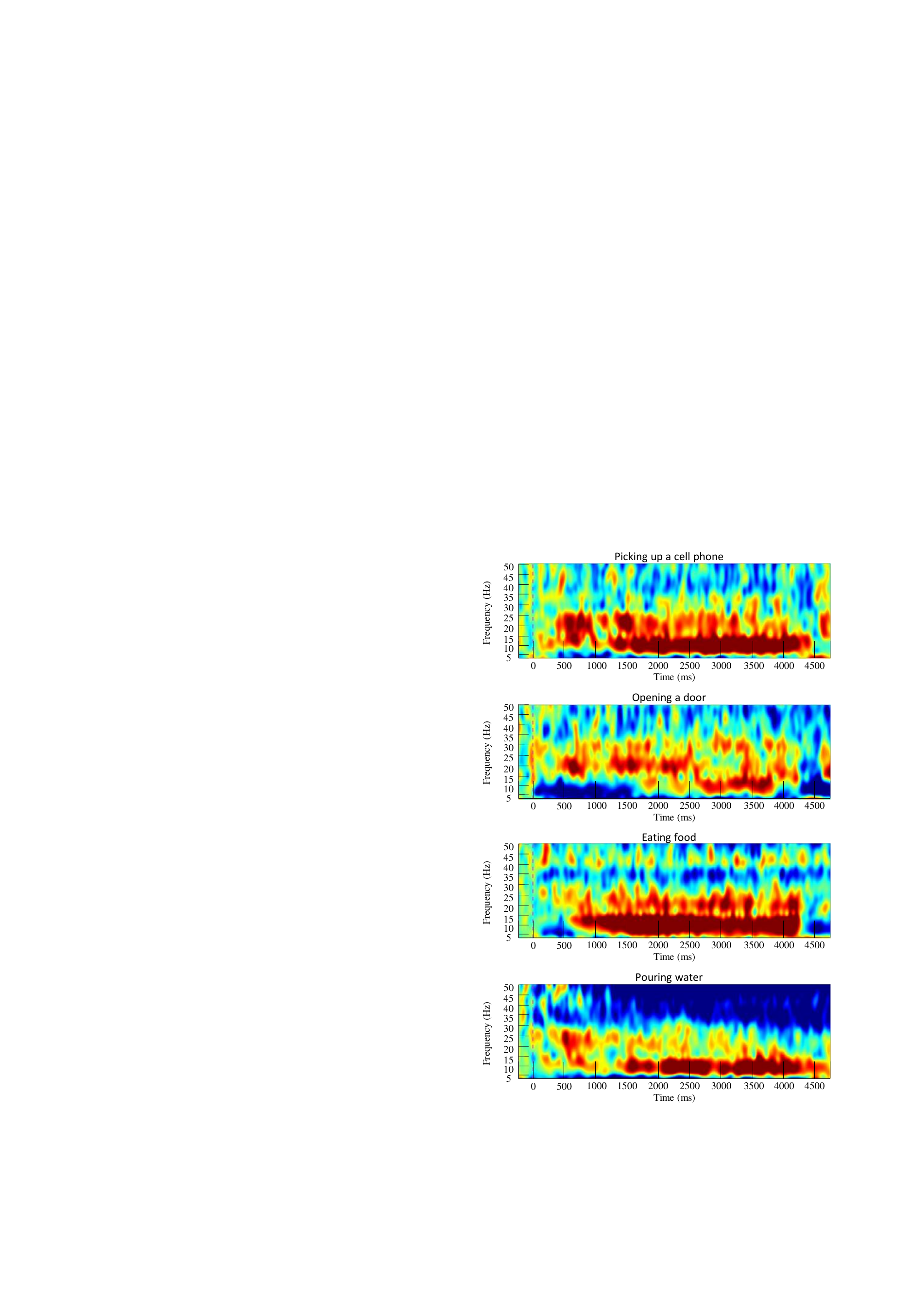}}
\caption{Event-related spatial perturbation of visual motion imagery in channel Oz. This indicates that the visual motion imagery shows significant activation between 0.5 and 13 Hz.}
\label{fig:5}
\end{figure}

\subsection{Performance Evaluation}
To verify the possibility of controlling the device with the visual motion imagery-based BCI, we validated our visual motion imagery data with various classification methods. We selected the most suitable number of channels based on functional connectivity and based on this selected channels, constructed an appropriate architecture. At this time, in addition to analyzing using 64 channels, the number of channels was selected as 2, 4, 8, 16, 20, and 32. The selected channels were selected in descending order of channel correlation based on functional connectivity.

 Table II shows the classification performance using the proposed method in each channel group. The performance with using 16 channels shows higher than using 64 channels. The results of comparing the architecture proposed in this study with CSP--LDA and deepConvNet can see in Table III. The performance when using 64 channels was 30.97\% for CSP--LDA, 63.47\% for deepConvNet, and 64.72\% for the proposed method. When using 16 channels, deepConvNet and the proposed method showed the highest accuracy, and the average of the results was 62.29\% and 67.5\%, respectively. Based on this result, it can be seen that the proposed architecture when using 16 channels shows high performance, which is higher than that of deepConvNet. This result indicates that by obtaining meaningful spatial information using channel selection based on functional connectivity, the spatial information determined to be noise is erased in advance, and based on this, temporal information can be obtained with appropriate depth architecture and it leads to high performance. This result supports that our hypothesis that spatial information can be obtained selectively through functional connectivity based channels and temporal information can be obtained through the proposed architecture.
 
\subsection{Neurophysiological Analysis}
Fig. 5 shows the variation of spectral power values for each frequency domain through the ERSP in the Oz channel, which is a channel in a region known to be significant in visual motion imagery. Each figure in Fig. 5 shows the result of the ERSP analysis of each class. Based on this, the significant features are shown in [0.5-13] Hz when performing visual motion imagery. According to Fig. 5, we could infer from the ERSP analysis results that it takes about 500 ms for the subject to start imagining. Therefore, in order to analyze with the significant feature of visual motion imagery, it is necessary to analyze [0.5-13] Hz after 500 ms as shown in Fig. 5. Also, based on the analysis of ERSP for each class, we could get information about whether or not subjects imagine each class, but we could not observe the significant features classified by class.

\section{Conclusion and Future Works}

In this study, we designed a virtual 3D BCI training platform for intuitive visual motion imagery (i.e., picking up a cell phone, opening a door, eating food, and pouring water). Moreover, we conducted a neurophysiological and statistical analysis to evaluate the framework and to find significant features in visual motion imagery. The results showed that the 3D-BCI training platform could induce visual motion imagery-related brain activities. Based on this, we decoded the intuitive visual motion imagery using functional connectivity based channel selection for spatial information and we proposed CNN architecture for decoding user intention robustly. We evaluated the classification performance using the conventional method. We confirmed 16 channels with the proposed architecture achieved the highest classification performance.

In future work, the proposed architecture using visual motion imagery will finally contribute to the development of BCI based devices such as a robotic arm driven by intuitive user intentions. Hence, we will overcome the lack of training set with various data augmentation methods and apply deeper neural network architecture to extract delicate temporal information. In addition, we plan to cover more multi-command for high degree of freedom. It can provide a convenient and high human and machine interaction effect for performing the high-level tasks.

\section*{Acknowledgment}
The authors thanks to Mr. G.-H. Shin for useful discussion of the data analysis and Mr. B.-H. Lee, and Ms. D.-Y. Lee for their help with the EEG database construction.\\


\bibliographystyle{IEEEtran}
\bibliography{references}

\begin{thebibliography}{10}
\providecommand{\url}[1]{#1}
\csname url@samestyle\endcsname
\providecommand{\newblock}{\relax}
\providecommand{\bibinfo}[2]{#2}
\providecommand{\BIBentrySTDinterwordspacing}{\spaceskip=0pt\relax}
\providecommand{\BIBentryALTinterwordstretchfactor}{4}
\providecommand{\BIBentryALTinterwordspacing}{\spaceskip=\fontdimen2\font plus
\BIBentryALTinterwordstretchfactor\fontdimen3\font minus
  \fontdimen4\font\relax}
\providecommand{\BIBforeignlanguage}[2]{{%
\expandafter\ifx\csname l@#1\endcsname\relax
\typeout{** WARNING: IEEEtran.bst: No hyphenation pattern has been}%
\typeout{** loaded for the language `#1'. Using the pattern for}%
\typeout{** the default language instead.}%
\else
\language=\csname l@#1\endcsname
\fi
#2}}
\providecommand{\BIBdecl}{\relax}
\BIBdecl

\bibitem{vaughan2003brain}
{T. M. Vaughan, W. J. Heetderks, L. J. Trejo, W. Z. Rymer, M. Weinrich, M. M.
  Moore, A. Kübler, B. H. Dobkin, N. Birbaumer, E. Donchin, E. W. Wolpaw, J.
  R. Wolpaw}, ``Brain-computer interface technology: {A} review of the second
  international meeting,'' \emph{IEEE Trans. Neural Syst. Rehabil. Eng.},
  vol.~11, no.~2, pp. 94--109, Jun. 2003.

\bibitem{abiri2018comprehensive}
{R. Abiri, S. Borhani, E. W. Sellers, Y. Jiang, and X. Zhao}, ``A comprehensive
  review of {EEG}-based brain-computer interface paradigms,'' \emph{J. Neural.
  Eng.}, vol.~16, no.~1, p. 011001, Jan. 2019.

\bibitem{zhu2016canonical}
{X. Zhu,H.-I. Suk, S.-W. Lee, and D. Shen}, ``{Canonical feature selection for
  joint regression and multi-class identification in Alzheimer’s disease
  diagnosis},'' \emph{Brain Imaging Behav.}, vol.~10, pp. 818--828, Aug. 2016.

\bibitem{ding2013changes}
{X. Ding and S.-W. Lee}, ``{Changes of functional and effective connectivity in
  smoking replenishment on deprived heavy smokers: a resting-state FMRI
  study},'' \emph{PLoS One}, vol.~8, p. e59331, Mar. 2013.

\bibitem{wolpaw2002brain}
{J. R. Wolpaw, N. Birbaumer, D. J. McFarland, G. Pfurtscheller, and T. M.
  Vaughan}, ``{Brain-computer interfaces for communication and control},''
  \emph{Clin. Neurophysio.}, vol. 113, pp. 767--791, Jun. 2002.

\bibitem{jeong2020decoding}
J.-H. Jeong, N.-S. Kwak, C.~Guan, and S.-W. Lee, ``Decoding movement-related
  cortical potentials based on subject-dependent and section-wise spectral
  filtering,'' \emph{IEEE Trans. Neural Syst. Rehabil. Eng.}, vol.~28, no.~3,
  pp. 687--698, Jan. 2020.

\bibitem{antelis2016decoding}
{J. M. Antelis, L. Montesano, A. Ramos-Murguialday, N. Birbaumer, J. Minguez},
  ``Decoding upper limb movement attempt from {EEG} measurements of the
  contralesional motor cortex in chronic stroke patients,'' \emph{IEEE Trans.
  Biomed. Eng.}, vol.~64, no.~1, pp. 99--111, Mar. 2016.

\bibitem{kwak2015lower}
N.-S. Kwak, K.-R. M{\"u}ller, and S.-W. Lee, ``A lower limb exoskeleton control
  system based on steady state visual evoked potentials,'' \emph{J. Neural
  Eng.}, vol.~12, no.~5, p. 056009, Aug. 2015.

\bibitem{benda2020comparison}
M.~Benda and I.~Volosyak, ``Comparison of different visual feedback methods for
  {SSVEP}-based {BCI}s,'' \emph{Brain Sci.}, vol.~10, no.~4, p. 240, Apr. 2020.

\bibitem{chen2016high}
Y.~Chen, A.~D. Atnafu, I.~Schlattner, W.~T. Weldtsadik, M.-C. Roh, H.~J. Kim,
  S.-W. Lee, B.~Blankertz, and S.~Fazli, ``A high-security {EEG}-based login
  system with {RSVP} stimuli and dry electrodes,'' \emph{IEEE Trans. Inf.
  Foren. Sec.}, vol.~11, no.~12, pp. 2635--2647, Jun. 2016.

\bibitem{fazel2012p300}
{R. Fazel, B. Z. Allison, C. Guger, E. W. Sellers, S. C. Kleih, and A.
  K{\"u}bler}, ``{P300 brain computer interface: current challenges and
  emerging trends},'' \emph{Front. Neuroeng}, vol.~5, p.~14, Jul. 2012.

\bibitem{li2018hybrid}
{J. Li, Z. L. Yu, Z. Gu, W. Wu, Y. Li, and L. Jin}, ``A hybrid network for
  {ERP} detection and analysis based on restricted {B}oltzmann machine,''
  \emph{IEEE Trans. Neural Syst. Rehabil. Eng.}, vol.~26, no.~3, pp. 563--572,
  Feb. 2018.

\bibitem{ang2017eeg}
{K. K. Ang, and C. Guan}, ``{EEG}-based strategies to detect motor imagery for
  control and rehabilitation,'' \emph{IEEE Trans. Neural Syst. Rehabil. Eng.},
  vol.~25, no.~4, pp. 392--401, Apr. 2017.

\bibitem{hochberg2012reach}
L.~R. Hochberg, D.~Bacher, B.~Jarosiewicz, N.~Y. Masse, J.~D. Simeral,
  J.~Vogel, S.~Haddadin, J.~Liu, S.~S. Cash, P.~Van Der~Smagt, and J.~P.
  Donoghue, ``Reach and grasp by people with tetraplegia using a neurally
  controlled robotic arm,'' \emph{Nature}, vol. 485, no. 7398, pp. 372--375,
  May 2012.

\bibitem{jeong2020Brain}
J.-H. Jeong, K.-H. Shim, D.-J. Kim, and S.-W. Lee, ``Brain-controlled robotic
  arm system based on multi-directional {CNN}-{BiLSTM} network using {EEG}
  signals,'' \emph{IEEE Trans. Neural Syst. Rehabil. Eng.}, Mar. 2020.

\bibitem{meng2018three}
J.~Meng, T.~Streitz, N.~Gulachek, D.~Suma, and B.~He, ``Three-dimensional
  brain--computer interface control through simultaneous overt spatial
  attentional and motor imagery tasks,'' \emph{IEEE Trans. Biomed. Eng.},
  vol.~65, no.~11, pp. 2417--2427, Oct. 2018.

\bibitem{kim2018commanding}
K.~T. Kim, H.-I. Suk, and S.-W. Lee, ``Commanding a brain-controlled wheelchair
  using steady-state somatosensory evoked potentials,'' \emph{IEEE Trans.
  Neural Syst. Rehabil. Eng.}, vol.~26, no.~3, pp. 654--665, Aug. 2018.

\bibitem{lee2015subject}
M.-H. Lee, S.~Fazli, J.~Mehnert, and S.-W. Lee, ``Subject-dependent
  classification for robust idle state detection using multi-modal neuroimaging
  and data-fusion techniques in {BCI},'' \emph{Pattern Recognit.}, vol.~48,
  no.~8, pp. 2725--2737, Aug. 2015.

\bibitem{edelman2019noninvasive}
{B. J. Edelman, J. Meng, D. Suma, C. Zurn, E. Nagarajan, B. S. Baxter, C. C.
  Cline, and B. He}, ``Noninvasive neuroimaging enhances continuous neural
  tracking for robotic device control,'' \emph{Sci. Robot.}, vol.~4, no.~31, p.
  eaaw6844, Jun. 2019.

\bibitem{jeong2020eeg}
J.-H. Jeong, B.-H. Lee, D.-H. Lee, Y.-D. Yun, and S.-W. Lee, ``E{EG}
  classification of forearm movement imagery using a hierarchical flow
  convolutional neural network,'' \emph{IEEE Access}, vol.~8, pp.
  66\,941--66\,950, Mar. 2020.

\bibitem{koizumi2018development}
{K. Koizumi, K. Ueda, and M. Nakao}, ``{Development of a cognitive
  brain-machine interface based on a visual imagery method},'' in \emph{Proc
  40th Annu. Int. Conf. of IEEE Engineering in Medicine and Biology Society
  (EMBC)}, 2018, pp. 1062--1065.

\bibitem{sousa2017pure}
{T. Sousa, C. Amaral, J. Andrade, G. Pires, U. J. Nunes, and M. B. Castelo},
  ``{Pure visual imagery as a potential approach to achieve three classes of
  control for implementation of BCI in non-motor disorders},'' \emph{J. Neural
  Eng.}, vol.~14, p. 046026, Jun. 2017.

\bibitem{blankertz2010berlin}
B.~Blankertz, M.~Tangermann, C.~Vidaurre, S.~Fazli, C.~Sannelli, S.~Haufe,
  C.~Maeder, L.~E. Ramsey, I.~Sturm, G.~Curio, and K.-R. M{\"u}ller, ``The
  {B}erlin brain-computer interface: non-medical uses of {BCI} technology,''
  \emph{Front. Neurosci.}, vol.~4, p. 198, Dec. 2010.

\bibitem{delorme2004eeglab}
A.~Delorme and S.~Makeig, ``{EEGLAB}: an open source toolbox for analysis of
  single-trial {EEG} dynamics including independent component analysis,''
  \emph{J. Neurosci. Methods}, vol. 134, no.~1, pp. 9--21, Mar. 2004.

\bibitem{maris2007nonparametric}
E.~Maris and R.~Oostenveld, ``Nonparametric statistical testing of {EEG}-and
  {MEG}-data,'' \emph{J. Neurosci. Methods.}, vol. 164, no.~1, pp. 177--190,
  Aug. 2007.

\bibitem{ang2012filter}
K.~K. Ang, Z.~Y. Chin, C.~Wang, C.~Guan, and H.~Zhang, ``Filter bank common
  spatial pattern algorithm on bci competition iv datasets 2a and 2b,''
  \emph{Front. Hum. Neurosci.}, vol.~6, p.~39, Mar. 2012.

\bibitem{schirrmeister2017deep}
{R. T. Schirrmeister, J. T. Springenberg, L. D. J. Fiederer, M. Glasstetter, K.
  Eggensperger, M. Tangermann, F. Hutter, W. Burgard, T. Ball}, ``Deep learning
  with convolutional neural networks for {EEG} decoding and visualization,''
  \emph{Hum. Brain Mapp.}, vol.~38, no.~11, pp. 5391--5420, Aug. 2017.

\end{thebibliography}

\end{document}